\documentclass[pra,aps,twocolumn,10pt,showpacs,groupedaddress,superscriptaddress,floatfix,notitlepage]{revtex4-2}
\usepackage{amsmath,amsfonts,amssymb,graphics,graphicx,epsfig,color,times}
\usepackage[utf8x]{inputenc}
\usepackage{color}
\usepackage{bbm, dsfont} 
\usepackage{subfigure}
\usepackage{hyperref}
\usepackage{mathrsfs}
\usepackage{verbatim}

\begin{document}
\title{Enhanced dark-state sideband cooling in trapped atoms via photon-mediated dipole-dipole interactions}

\author{Chung-Hsien Wang}
\email{b07202032@ntu.edu.tw}
\affiliation{Department of Physics, National Taiwan University, Taipei 10617, Taiwan}
\affiliation{Institute of Atomic and Molecular Sciences, Academia Sinica, Taipei 10617, Taiwan}

\author{Yi-Cheng Wang}
\email{r09222006@ntu.edu.tw}
\affiliation{Department of Physics, National Taiwan University, Taipei 10617, Taiwan}
\affiliation{Institute of Atomic and Molecular Sciences, Academia Sinica, Taipei 10617, Taiwan}

\author{Chi-Chih Chen}
\affiliation{Institute of Atomic and Molecular Sciences, Academia Sinica, Taipei 10617, Taiwan}
\affiliation{Department of Electrical and Computer Engineering, Stony Brook University, Stony Brook, NY 11794, USA}

\author{Chun-Che Wang}
\affiliation{Institute of Atomic and Molecular Sciences, Academia Sinica, Taipei 10617, Taiwan}
\affiliation{Department of Physics and Astronomy, University of Pittsburgh, Pittsburgh, PA 15260, USA}

\author{H. H. Jen}
\email{sappyjen@gmail.com}
\affiliation{Institute of Atomic and Molecular Sciences, Academia Sinica, Taipei 10617, Taiwan}
\affiliation{Physics Division, National Center for Theoretical Sciences, Taipei 10617, Taiwan}

\date{\today}
\renewcommand{\r}{\mathbf{r}}
\newcommand{\f}{\mathbf{f}}
\renewcommand{\k}{\mathbf{k}}
\def\p{\mathbf{p}}
\def\q{\mathbf{q}}
\def\bea{\begin{eqnarray}}
\def\eea{\end{eqnarray}}
\def\ba{\begin{array}}
\def\ea{\end{array}}
\def\bdm{\begin{displaymath}}
\def\edm{\end{displaymath}}
\def\red{\color{red}}
\pacs{}
\begin{abstract}
Resolved sideband cooling provides a crucial step in subrecoil cooling the trapped atoms toward their motional ground state, which is essential in atom-based quantum technologies. Here we present an enhanced dark-state sideband cooling in trapped atoms utilizing photon-mediated dipole-dipole interactions among them. By placing the atoms at the magic interparticle distances, we manifest an outperformed cooling behavior in the target atom, which surpasses the limit that a single atom permits. We further investigate various atomic configurations in a multiatom setup with a laser detuning and different light polarization angles, where multiple magic spacings can be identified and a moderate improvement in cooling performance is predicted as the number of atoms increases. Our results provide insights to subrecoil cooling of atoms with collective and light-induced long-range dipole-dipole interactions, and pave the way toward implementing genuine quantum operations in multiple quantum registers.
\end{abstract}
\maketitle

\section{Introduction} 

Laser-cooling technique in trapped ions \cite{Leibfried2003} or atoms \cite{Cohen2011} has been an essential and indispensable element in many quantum technology applications such as quantum simulations \cite{Buluta2009, Lanyon2011, Monroe2021, Bernien2017, Ebadi2021, Scholl2021} and quantum computation \cite{Cirac1995, Kielpinski2002, Pino2021, Bluvstein2022, Graham2022}. To further cool the system into subrecoil cooling regime, a resolved sideband cooling \cite{Diedrich1989, Cirac1992, Monroe1995, Roos1999, Kaufman2012, Zhang2021} or dark-state sideband cooling utilizing electromagnetic induced transparency (EIT) \cite{Morigi2000, Roos2000, Morigi2003, Cerrillo2010, Kampschulte2014, Lechner2016, Jordan2019, Feng2020, Qiao2021} in the Lamb-Dicke (LD) regime provides a recipe to approach its motional ground state. This nearly zero-phonon state can be prepared owing to the red-sideband transition which drives the trapped atoms between their ground and excited states with one phononic quanta less, and relaxes to a ground state via spontaneous emissions with one phonon removed effectively. In this resonantly-driven sideband transition, the carrier and blue-sideband transitions that cause heating can be suppressed. This leads to a lower bound in the phonon occupation, which is determined by the system's finite spontaneous emission rate.   

In a multiple atomic system, the cooling mechanism becomes more intriguing in a sense that spin-phonon correlations can arise via light-matter interactions \cite{Jordan2019, Shankar2019}. This correlation builds up via photon-mediated dipole-dipole interactions (DDIs) in free space \cite{Lehmberg1970, Cirac2000, Harlander2011, Meir2014, Rui2020, Wang2022_mirror}, which can further be tailored into a strong coupling regime in an atom-waveguide interface \cite{Chang2018, Corzo2019, Sheremet2022} with controllable directionality of the spin-exchange coupling \cite{Mitsch2014, Lodahl2017, Albrecht2019, Jen2020_PRR, Jen2021_bound, Zanner2022, Pennetta2022, Jen2022_correlation, Pennetta2022_2}. Using this photon-mediated and collective DDIs as cooling mechanism has been investigated in two ions in free space \cite{Vogt1996}, one-dimensional ion crystal mediating with guided modes \cite{Chen2022}, cavity-mediated atoms \cite{Xu2016}, optically bound cold atoms \cite{Maximo2018, Gisbert2019}, and two atoms in an atom-waveguide interface \cite{Wang2022}. This presents a pathway, not a shortcut, to a superior cooling behavior than in a single atomic system via collective spin-exchange interactions between constituent atoms.  

In a chirally-coupled atom-waveguide interface, a superior cooling behavior emerges from nonreciprocal couplings that allow distinct and directional heat removal, where the steady-state phonon occupation of the target atom under asymmetric driving conditions can surpass the single atom limit toward a cooler regime \cite{Chen2022, Wang2022}. By contrast, a heating behavior arises at the reciprocal coupling regime owing to strong spin-spin correlations among the atoms \cite{Chen2022}. Interestingly, this heating mechanism can be modified and turned into cooling instead by introducing nonguided modes in the interface. This shows a more realistic scenario in most of the atom-waveguide systems, where finite nonguided modes in the system give rise to a less strongly-coupled regime, while these extra nonguided modes can reduce the spin-spin correlations relevant in heating and lead to a new parameter region that allows cooling. We find that a similar form of photon-mediated DDIs in free space resembles the ones in the atom-waveguide interface at reciprocal couplings with nonguided modes. This guides us further in this paper to explore the possibilities for better cooling behaviors in free-space trapped atoms with assistance of photon-mediated DDIs, which provides a surpassing mechanism to go around the cooling barriers encountered in noninteracting single atoms. 

\begin{figure}[t]
    \includegraphics[width=0.45\textwidth]{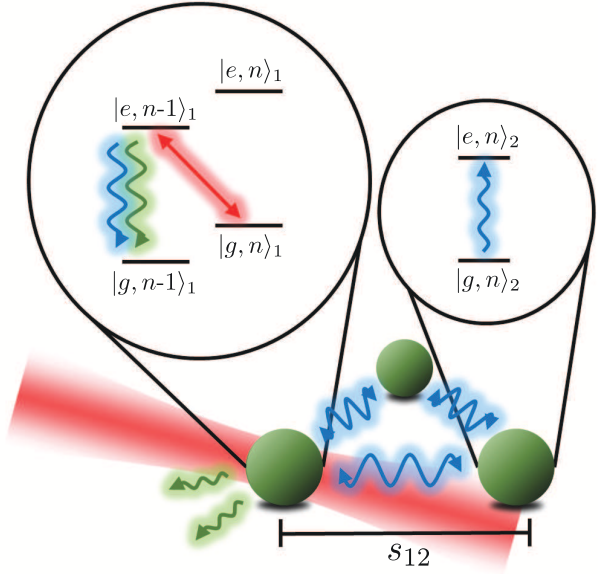}
    \caption{Schematic plot of sideband cooling under photon-mediated DDIs on an equilateral triangle with an interparticle distance $s_{12}$. A laser (red beam and arrow) is applied on the target atom and resonantly drives the red-sideband transition $|g,n\rangle \to |e,n-1\rangle$, which cools the atom toward $|g,n-1\rangle$ through spontaneous emission (green arrow). The effect of photon-mediated DDIs manifests in the spin-exchange interactions (blue arrow), which spin flips the other atoms through multiple light scattering with a conservation of total atomic excitations. The other two spectator atoms serve as refrigerants and provide a distinct cooling mechanism of removing extra heat from the target atom via collective and resonant DDIs.}\label{fig1}
\end{figure}

Here we employ the dark-state sideband cooling scheme in trapped atoms as shown in Fig. \ref{fig1}, where an effective two-level structure of atoms can be denoted as $|e\rangle$ and $|g\rangle$, which can be constructed by two hyperfine ground states coupled by two laser fields in a $\Lambda$-type atomic structure. The photon-mediated DDIs emerge owing to rescattering events of light mediating all constituent atoms. Under this collective spin-exchange interactions and assuming asymmetric driving conditions, we present a surpassing cooling performance from the steady-state phonon occupations in the target atom by placing the atoms at the magic interparticle distances. We further investigate different atomic configurations in a multiatom setup with a laser detuning and different light polarization angles, where we find more magic spacings that allow superior cooling behaviors and demonstrate the multiatom improvement in cooling as the number of atoms increases. 

The remainder of the paper is organized as follows. In Sec. II, we introduce the Hamiltonian and master equation with Lindblad forms of photon-mediated DDIs. In Sec. III, we define the magic interparticle distances by locating the zeros of the energy shifts in the photon-mediated DDIs. We further explore various atomic configurations in a two-dimensional space, including the effect of laser detuning, light polarization angles, collective frequency shifts, and the number of atoms, on the cooling performance of the target atom. Finally, we discuss and conclude in Sec. IV.   
  

\section{Hamiltonian and Master Equation}

We consider a conventional model of $N$ trapped atoms under the standing-wave sideband cooling with a two-level atomic structure \cite{Cirac1992, Zhang2021, Chen2022} or the dark-state sideband cooling within two hyperfine ground states using a $\Lambda$-type atomic configuration \cite{Morigi2000, Wang2022}. In this two-level atomic structure, all atoms are confined locally in a one-dimensional harmonic trap potential with a frequency $\nu$, which should be the most weakly-confined direction compared to the other spatial degrees of freedom. In the dark-state sideband cooling scheme, the effective two-level structure can be formed by quantum interference through two laser fields operating on one common excited state. Under the condition of large two-photon detuning, a manifold of two hyperfine ground states construct the dark-state bases, resembling the sideband cooling scheme but with a modified laser coupling strength and an associated spontaneous emission rate \cite{Zhang2021, Wang2022}. Here we use $|g\rangle$ and $|e\rangle$ as an effective two-level atomic structure in Fig. \ref{fig1}, which can equivalently be mapped to the dark-state bases, and we note that these system parameters can be tunable and tailored to ensure the sideband cooling condition. 

In the LD limit where its parameter $\eta\equiv k_{\rm eff}/\sqrt{2m\nu}\ll 1$ with an atomic mass $m$ and an effective wave vector $k_{\rm eff}$, we have
\begin{eqnarray}
    H_{\rm LD}=&& -\Delta \sum_{\mu=1}^N \sigma_\mu^\dagger \sigma_\mu + \nu \sum_{\mu=1}^N a_\mu^\dagger a_\mu  \nonumber \\&&+ \frac{\eta}{2} \sum_{\mu=1}^N \Omega_\mu (\sigma_\mu^\dagger + \sigma_\mu)(a_\mu^\dagger + a_\mu),
\end{eqnarray}
where $\sigma_\mu=|g\rangle_\mu\langle e|$ ($\sigma_\mu^\dag$) represents the lowering (raising) operator and $a_\mu$ ($a_\mu^\dag$) is the creation (annihilation) operator in a phononic Fock space $|n\rangle$. The laser Rabi frequency is denoted by $\Omega_\mu$ with an atomic label $\mu$ for inhomogeneous driving conditions, and a detuning $\Delta=\omega_L-\omega_{eg}$ indicates the difference between the laser central frequency and the atomic transition frequency in the sideband cooling scheme \cite{Cirac1992} or an eigen-energy shift $\Delta=-\omega_e$ for the eigenstate $|e\rangle$ in the dark-state basis \cite{Morigi2000} where $\omega_g=0$ for $|g\rangle$. The $k_{\rm eff}$ can be $\omega_{eg}/c$ or ($k_1\cos\phi_1-k_2\cos\phi_2$) in the dark-state basis with two laser wave numbers $k_{1(2)}$ and corresponding projection angles $\phi_{1(2)}$ to the motional direction. 

The system dynamics can then be described by a master equation with a density matrix $\rho$ for $N$ atoms, which reads  
\begin{eqnarray}
    \dot{\rho} =&&   -i[H_{\rm LD},\rho] + \sum_{\mu=1}^N \Gamma \mathcal{L}_{\mu\mu}[\rho]  -i\sum_{\mu \ne \nu}^N\sum_{\nu=1}^N g_{\mu\nu}[\sigma_\mu^\dagger\sigma_\nu,\rho] \nonumber \\ &&+ \sum_{\mu \ne \nu}^N \sum_{\nu=1}^N \gamma_{\mu\nu} \mathcal{L}_{\mu\nu}[\rho],\label{rho}
\end{eqnarray}
where the Lindblad map is defined by $\mathcal{L}_{\mu\nu}[\rho] = \sigma_\nu\rho \sigma_\mu^\dagger -\frac{1}{2}\{\sigma_\mu^\dagger\sigma_\nu,\rho\}$ with
\begin{eqnarray}
    g_{\mu\nu} =&&  \frac{3\Gamma}{4} \bigg\{ -\left[1-\cos^2 \theta\right]\frac{\cos \xi}{\xi} \nonumber \\&&+    \left[1-3\cos^2 \theta\right]\left[ \frac{\sin \xi}{\xi^2} + \frac{\cos \xi}{\xi^3} \right] \bigg\},\label{g}\\
    \gamma_{\mu\nu} =&& \frac{3\Gamma}{2} \bigg\{ \left[1-\cos^2 \theta\right]\frac{\sin\xi}{\xi} \nonumber \\&&+ \left[1-3\cos^2 \theta\right]\left[ \frac{\cos\xi}{\xi^2} - \frac{\sin\xi}{\xi^3} \right] \bigg\}.\label{r}
\end{eqnarray}
The effect of photon-mediated DDIs \cite{Lehmberg1970} manifests in $g_{\mu\nu}$ and $\gamma_{\mu\nu}$, which represent the collective frequency shifts and decay rates, respectively, with $\Gamma$ quantifying the effective decay rate from $|e\rangle$. They depend on the relative distances between atoms, $\xi = k|\vec{s}_{\mu\nu}| \equiv k|\vec{r}_\mu-\vec{r}_\nu|$, showing the long-range nature of pairwise interactions \cite{Jen2022_EIT}, and also their dipole polarization angle $\cos \theta \equiv \hat{ \textbf{d}}\cdot \hat{s}_{\mu\nu}$, where $\hat{ \textbf{d}}$ is the unit vector of polarization. In general, $\rho$ involves a tensor product of spin and phononic states for all atoms, $\prod_{\mu=1}^N\sum_{n=0}^{n_c} |g(e),n\rangle_\mu$, where we introduce a cutoff number of Fock states $n_c$ in practice, which should be large enough for convergence in steady-state calculations. This is particularly valid in the sideband cooling regime with a finite $n_c$, where $\langle n_\mu \rangle = \text{tr}(a_\mu^\dagger a_\mu \rho_{\rm st})\ll 1$ under a steady-state density matrix $\rho_{\rm st}$ from $\dot\rho_{\rm st}=0$ in Eq. (\ref{rho}). 

The essence of sideband cooling mechanism results from a resolved and red-sideband transition, $\nu\gg \eta\Omega_\mu,\Gamma$ and $\Delta = -\nu$, which decouples the atoms from blue-sideband and carrier transitions, and transfers the atom toward the motional ground state as $\langle n_1\rangle\propto (\Gamma/\nu)^2$ \cite{Cirac1992}. This lowest possible phonon occupation is limited by the spontaneous emission strength $\Gamma$, which can be broken by introducing nonreciprocal and collective decay channels in a one-dimensional resonant DDIs of an atom-waveguide interface, where a minimal value, as $\eta\Omega_1\rightarrow 0$, can reach without limit as $\langle n_1\rangle^{\rm min}\propto (\eta\Omega_1\Gamma/\nu^2)$ \cite{Chen2022, Wang2022}. This shows a new realm of using collective DDIs as a novel cooling mechanism in nanophotonic devices to go beyond the single atom limit in free space. Below we further explore the enhanced cooling scheme using photon-mediated DDIs in free space by tailoring various atomic configurations. 

\section{Enhanced dark-state sideband cooling}

In this section we present our main results of enhanced cooling behaviors utilizing photon-mediated DDIs. We first show the magic spacing in few-atom configurations, which manifests superior cooling performance. We further elaborate more on the other magic interparticle distances that also allow enhanced cooling owing to the variations of light polarization orientations. Then we continue the investigation on the laser detuning and multiatom configurations. Essentially, up to $13\%$ to $17\%$ reduction of the phonon occupations compared to the single atom case can be reached in our considerations. This shows the capability of photon-mediated DDIs in further cooling of free-space atoms beyond the limit in a single atom setup. 

\subsection{Two and three atoms cases}

Here we first consider two- and three-atom configurations, where one of them is denoted as the target atom and the rest of them are spectator or refrigerant atoms. As shown in Fig. \ref{fig1}, the role of the spectator atoms serves as quantum emitters to mediate collective spin-exchange interactions. Under the asymmetric setting where only the target atom is driven by the laser field, we are able to neglect the motional degrees of the residual atoms (corrections of higher order than $\eta^2\Omega_1^2/\nu^2$ in phonon occupations) and keep the target atom's phononic degrees of freedom, which hugely reduces the complete and many-body spin-phonon spaces in Eq. (\ref{rho}). This asymmetric driving condition has been investigated in sideband cooling of ions and EIT cooling of neutral atoms in an atom-waveguide interface \cite{Chen2022, Wang2022}, which allows new parameter regimes for superior cooling behaviors than the single atom case.  

\begin{figure}[b]
    \includegraphics[width=0.48\textwidth]{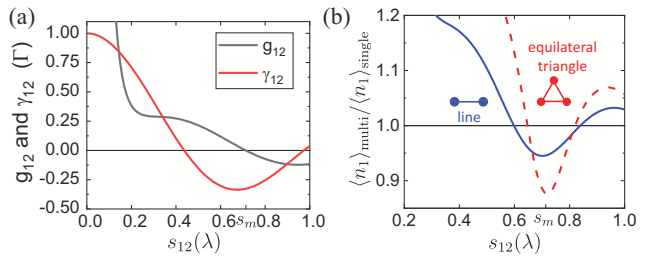}
    \caption{Magic spacing and enhanced sideband cooling in few-atom cases. (a) Collective energy shift $g_{12}$ and decay rate $\gamma_{12}$ with a perpendicular polarization for various atomic separation $s_{12}$. A magic spacing $s_m \simeq 0.7133 \lambda$ can be identified where $g_{12}$ vanishes. (b) A normalized phonon occupation of the target atom $\langle n_1 \rangle_{\text{multi}}$ over the single atom result $\langle n_1 \rangle_{\text{single}}$ in the multiatom setups of atoms forming a line (blue-solid line) or an equilateral triangle (red-dash line) for $s_{12}$ within $\lambda$. The sideband cooling condition is chosen that $\Delta = -\nu$, $\Gamma = 0.1\nu$, and $\eta\Omega_1 = 0.04\nu$. A maximal superior cooling appears when $s_{12}$ is chosen at the magic spacing $s_m$.}\label{fig2}
\end{figure}

\begin{figure*}[t]
    \includegraphics[width=0.96\textwidth]{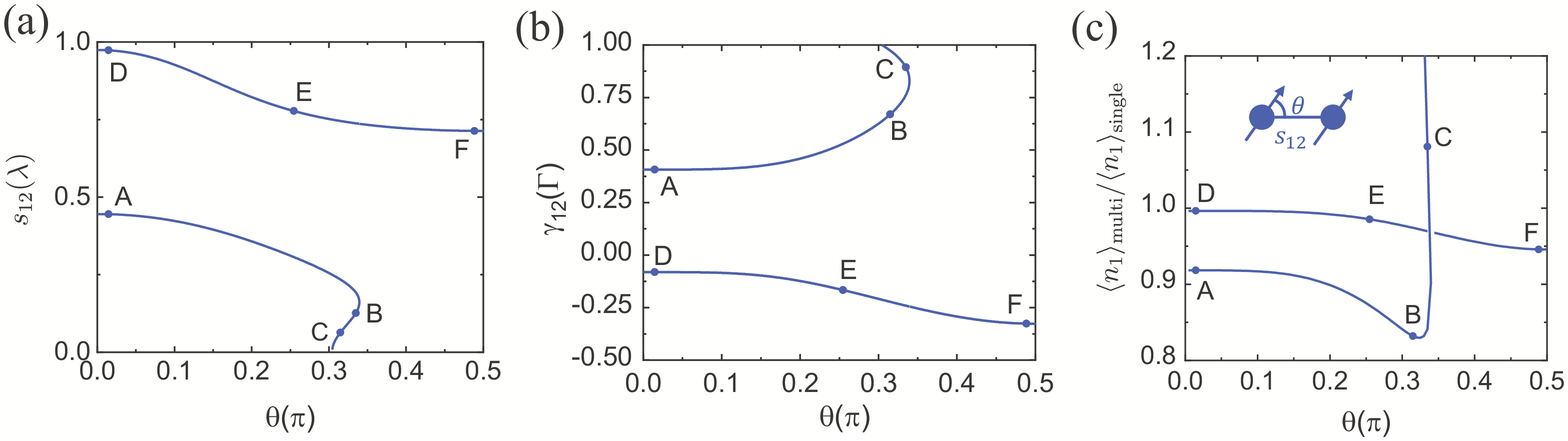}
    
    \caption{Magic spacings, associated collective decay rate, and enhanced cooling in a two-atom setup. (a) A set of $(s_{12},\theta)$ denoted by solid lines with vanishing frequency shifts $g_{12}(s_{12},\theta)$ with $\cos \theta \equiv {\bf \hat{d}} \cdot \hat{s}_{12}$ and $s_{12}\leq\lambda$. We select A to F as representative points of magic spacings, at which the corresponding collective decay rates $\gamma_{12}$ and normalized cooling behaviors are shown in (b) and (c), respectively. The definition of $\theta$ can be seen clearly in the inset plot of (c), and multiple magic spacings show up at $g_{12}=0$ when $\theta$ is varied. Other system parameters are same as those in Fig. \ref{fig2}.}\label{fig3}
\end{figure*}

We numerically obtain the steady-state solutions of the target atom $\langle n_1 \rangle_{\text{multi}}$ in a multiatom setup with photon-mediated DDIs from Eq. (\ref{rho}) by choosing $n_c=1$. As a comparison, we normalize the results by the single atom value $\langle n_1 \rangle_{\text{single}}$ without photon-mediated DDIs, where the ratio $\langle n_1 \rangle_{\text{multi}}/\langle n_1 \rangle_{\text{single}}$ shows enhanced cooling behaviors when it is below one. We can identify the magic spacing in the two-atom case in Fig. \ref{fig2}(a), which corresponds to a maximally enhanced performance of sideband cooling in Fig. \ref{fig2}(b). Under the perpendicular polarization case where $\hat{ \textbf{d}}\cdot \hat{s}_{\mu\nu} = 0$, we locate the interparticle spacing $s_m\approx 0.71\lambda$ as a magic distance where its corresponding collective frequency shift $g_{12}$ vanishes. This absence of collective frequency shift resembles the cooling condition at $\xi=0,\pi$ in $\exp(i\xi)$ \cite{Chen2022, Wang2022} as the spatial dependence of an infinite-range photon-mediated DDIs in a one-dimensional atom-waveguide platform \cite{Mitsch2014,Lodahl2017}. At $s_m$ with a finite $\gamma_{12}$ and $|\gamma_{12}|<\Gamma$ (as it is always true for $\xi\neq 0$) as shown in Fig. \ref{fig2}(a), this condition accesses the cooling regimes as in an atom-waveguide system at reciprocal couplings with nonguided modes which can be quantified exactly as $(\Gamma-|\gamma_{12}|)$ \cite{Chen2022}. 

In Fig. \ref{fig2}(b), a configuration of equilateral triangle with an interparticle distance at the magic spacing, $s_{\mu\nu}=s_m$, shows an example of extension to the two-atom case. All mutual energy shifts and decay rates are equivalent to the case of two atoms in a line, and $g_{\mu\nu}$ vanishes simultaneously in these configurations, near which a maximal cooling performance emerges. A finite region of enhanced cooling performances can be seen as well for $s_{12}$ around $s_m$, where $|g_{12}|$ arises and compromises the cooling performance. Later in Sec. III. C, we will show that this effect can be counteracted by a laser detuning, which leads to new cooling parameter regions and brings the system back to enhanced cooling again. In the following discussion, all the polarizations $\hat{ \textbf{d}}$ are set in the $z$ direction and the atoms are placed on the $x$-$y$ plane without loss of generality. Throughout the paper we also fix the value of $\eta\Omega_1$ as in Fig. \ref{fig2}(b) in general. We note that the performance of cooling can further be improved as $\eta\Omega_1$ decreases, but its cooling rate would be compromised owing to weak laser fields.

\subsection{Various magic spacings}

Here we further explore other possible magic spacings in a two-atom setup as we vary the polarization angles $\theta$. In Eq. (\ref{g}), the polarization angle modifies the short-range behaviors of the collective frequency shifts, and more crossing points arise at $g_{12}=0$ when $\theta$ is varied. In Fig. \ref{fig3}(a), we show these crossing points with $s_{12}\leq \lambda$, where we identify the magic spacings from $g_{12}(s_{12},\theta)=0$. As $\theta$ is tuned toward zero, a parallel polarization to the line axis of two atoms, a multiple of magic spacings show up and allow possible enhanced cooling behaviors. We choose six representative points from A to F to trace the changes of the associated collective decay rates $\gamma_{12}$ and corresponding cooling performances in Figs. \ref{fig3}(b) and \ref{fig3}(c), respectively. For $s_{12}> \lambda$, $\gamma_{12}$ in Eq. (\ref{r}) decays as $\propto 1/\xi$ at long distances, which becomes weakened, turning the system toward a noninteracting regime with less significant photon-mediated DDIs. The point F at $\theta\approx\pi/2$ corresponds to the case in Fig. \ref{fig2}, where the cooling performance is moderate as $\langle n_1\rangle_{\rm multi}/\langle n_1\rangle_{\rm single}\approx 0.95$, a $5\%$ reduction from the single atom case. 

In Figs. \ref{fig3}(b) and \ref{fig3}(c), we can trace the changes of $\gamma_{12}$ and cooling behaviors according to the magic spacings in Fig. \ref{fig3}(a). The positive and negative $\gamma_{12}$ corresponds to the set of magic spacings $s_{12}=s_m$ determined by the crossing points when $g_{12}=0$ and demonstrates multiple oscillations of $g_{12}$ within an interparticle distance of $\lambda$ when $\theta\lesssim 0.35\pi$. Within this interparticle range, we find that the maximal cooling performance appears around the point B, which is optimal with an approximately $17\%$ reduction compared to the single atom result. In its trace between the points A and C, we find significant $|\gamma_{12}|/\Gamma\gtrsim 0.5$ comparing the corresponding values in the trace D-E-F. This shows that an optimal cooling condition emerges at a finite $|\gamma_{12}|$ satisfying $0.5<|\gamma_{12}|/\Gamma<0.8$, in between a moderately interacting regime where $\gamma_{12}/\Gamma\approx 0.4$ (point A) and the heating regime where $\gamma_{12}\rightarrow \Gamma$ (point C). In the moderately interacting regime, photon-mediated DDIs play some role in initiating the cooling mechanism to remove the extra heat with assistance of a finite $(\Gamma-|\gamma_{12}|)$ as the nonguided modes in an atom-waveguide system. On the other hand, near the point C, the magic spacing is very close to $0.05\lambda$, reaching the strongly interacting regime that hosts Dicke superradiance \cite{Dicke1954} with high spin-spin correlations, and thus leads to heating instead. This particular heating mechanism owing to strong spin-spin correlations is also predicted in an atom-waveguide interface specifically at reciprocal couplings without nonguided modes ($\Gamma-|\gamma_{12}|\rightarrow 0$) \cite{Chen2022}, which retains the phonons within the constituent atoms and effectively heats up the system.

\begin{figure*}[ht]
    \includegraphics[width=0.96\textwidth]{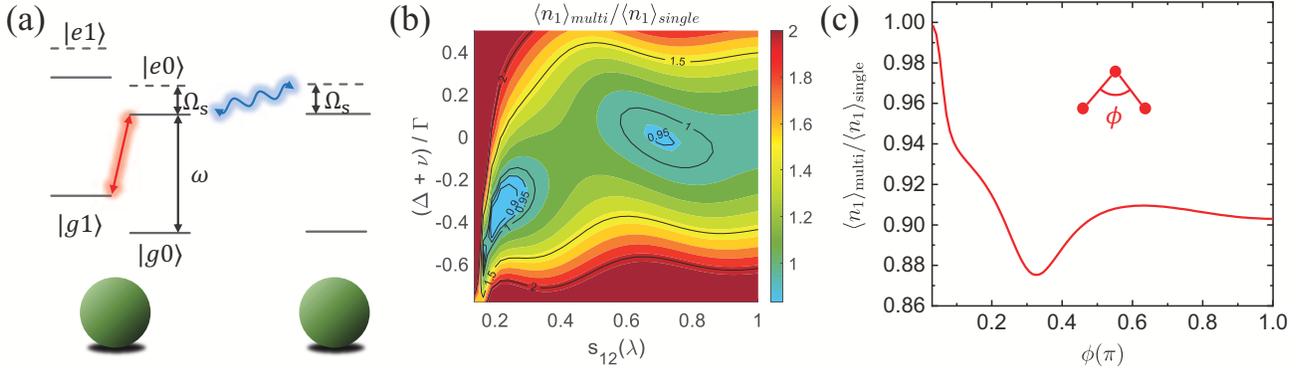}  
    \caption{The effect of laser detuning and collective frequency shift on the cooling performance. (a) A schematic plot shows the effect of nonvanishing collective frequency shift $\Omega_s$ in the spin-exchange processes (blue arrow), which leads to a worse cooling behavior when the laser (red arrow) is tuned on the conventional red-sideband cooling condition. (b) A diagram of cooling performance for laser detuning and interparticle distances in the two-atom case. (c) An isosceles triangle configuration with equal side lengths at $s_m\approx 0.71\lambda$ presents the optimal enhanced cooling at $\phi = \pi/3$ on the target atom at the top vertex, where the best cooling behavior emerges with vanishing collective frequency shifts for the target and the spectator atoms. For other angles $\phi$ corresponding to different three-atom configurations (say a line of atoms at $\phi=\pi$), a suppressed exchange process owing to finite $\Omega_s$ leads to a worse performance but still better than the single atom case. Other system parameters are same as those in Fig. \ref{fig2}.}\label{fig4}
\end{figure*}

In the trace of the points D-E-F, they are close to the noninteracting regime where $|\gamma_{12}|\rightarrow 0$. In the noninteracting regime, photon-mediated DDIs become irrelevant, and the cooling performance we defined approaches the single atom result, that is, no significant enhanced cooling can be identified. This exactly reflects to the result at the point D where $\langle n_1\rangle_{\rm multi}/\langle n_1\rangle_{\rm single}\approx 1.0$. Comparing the points A and F, $|\gamma_{12}|$ at A is slightly larger than the one at F, which results in a slightly more enhanced cooling performance at A as shown in Fig. \ref{fig3}(c). 

\subsection{Effect of laser detuning and collective frequency shift}

Next, we release the exact sideband cooling condition of $\Delta=-\nu$ and investigate the role of laser detuning with non-vanishing collective frequency shifts in the cooling performance of constituent systems. As shown schematically in Fig. \ref{fig4}(a), as the interparticle distance $s_{12}$ varies, a finite $\Omega_s=g_{12}$ arises when $s_{12}\neq s_m$. This finite frequency shift degrades the best cooling performance, but can be compensated by the driving field detuning. We expect that a detuned laser field away from the sideband cooling condition would allow new parameter regimes of cooling. 

In Fig, \ref{fig4}(b), we numerically calculate the cooling performance with dependence of laser detuning and interparticle distances in the setup of two atoms. In this plot, we can see that the magic spacing emerges at the value we have found in Fig. \ref{fig2} when $\Delta=-\nu$. When this condition is released, we obtain another magic spacing at $s_{12}/\lambda\approx 0.25$ when $(\Delta+\nu)/\Gamma\approx-0.3$, which corresponds to $g_{12}/\Gamma\approx 0.3$ as shown Fig. \ref{fig2}(a). This presents that the frequency shift can be counteracted by the laser detuning, and the cooling behavior at this second magic spacing shows an even better performance than the one we have located in Fig. \ref{fig2}(b). We note that a banded area of magic spacings for significant enhanced cooling is not spotted here, but only two restricted areas are shown for superior cooling performance. Since the cooling performance is determined by both $g_{12}$ and $\gamma_{12}$ simultaneously, it is not necessarily that an enhanced cooling appears at $\Delta+\nu\approx -g_{12}$ without significant $|\gamma_{12}|$. The area for $\langle n_1\rangle_{\rm multi}/\langle n_1\rangle_{\rm single}\approx 1.0$, however, follows closely the condition of $\Delta+\nu\approx -g_{12}$, and encloses these two restricted areas. At $s_{12}/\lambda\leq 0.2$, a shrinking and narrow region of enhanced cooling reflects the abrupt rise of $g_{12}$ which requires an even larger red detuning of the laser field to counteract its effect before the heating mechanism takes over as $\gamma_{12}$ approaches $\Gamma$. Similar heating effect can be seen as well at the point C in Figs. \ref{fig3}(b) and \ref{fig3}(c). 

In Fig. \ref{fig4}(c), we consider an isosceles triangle configuration, where the target atom is placed at the top vertex of this triangle with equal side lengths of $s_{m}$ for vanishing collective frequency shifts. As the angle $\phi$ changes, a finite and varying collective frequency shifts on the refrigerant atoms can arise and thus modify the cooling behavior in the target atom. The best enhanced cooling performance emerges at $\phi=\pi/3$, which combining the result of Fig. \ref{fig2}(b) indicates that the optimal cooling behavior favors an equilateral triangle structure in the $\phi$-$s_{12}$ phase spaces. This shows the effect of frequency shift either on the target and the spectator atoms in Fig. \ref{fig2}(b) or only on the spectator atoms in Fig. \ref{fig4}(c), where spin-exchange-assisted enhanced cooling can be allowed in constituent atoms with photon-mediated DDIs. An optimal of $\langle n_1\rangle_{\rm multi}/\langle n_1\rangle_{\rm single}\approx 0.87$ can be obtained with an enhanced cooling in broad parameter regimes of $\phi$, unless when $\phi\rightarrow 0$, reaching the cooling behavior of a single atom owing to the divergent frequency shift between the spectator atoms. This indicates the suppression of collective spin-exchange interactions owing to nonresonant light-atom couplings and leads to an ineffective cooling mechanism.  

\subsection{Multiatom configurations}

Finally, we explore multiatom configurations based on the previous findings. As previous sections indicate, the collective frequency shifts on the target atom should be avoided under the sideband cooling condition to allow for the best enhanced cooling performance. Therefore, we put the target atom at the center of a hexagonal structure with the other spectator atoms placed on its vertices and consider the configurations up to five atoms. By choosing the side lengths equivalent to the magic spacing, the target atom experiences null frequency shifts of photon-mediated DDIs from the other atoms. This way we are able to investigate the structure and multiatom effect on the cooling performance of the target atom. 

\begin{figure}[tb]
     \includegraphics[width=0.46\textwidth]{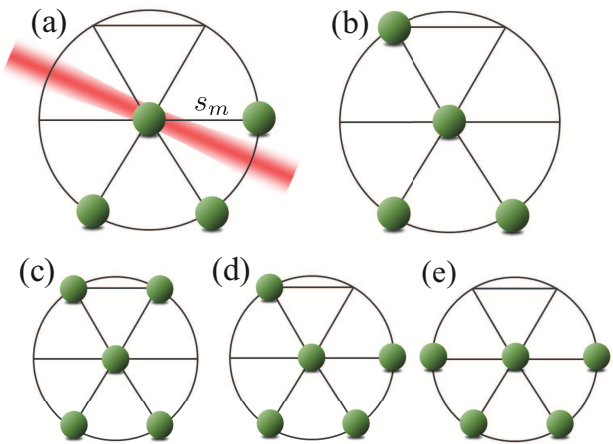}
    \caption{All possible (a-b) $4$- and (c-e) $5$-atom configurations placed at the center and on the hexagon vertices. The target atom lies at the center with hexagon's side lengths chosen as magic spacings $s_m\approx0.71\lambda$. The cooling enhancement can be obtained from $\langle n_1 \rangle_{\rm multi} / \langle n_1 \rangle_{\rm single}$ which are $0.863$, $0.864$, $0.842$, $0.849$ and $0.890$, respectively, in (a-e), showing a moderate enhancement in cooling behaviors comparing the cases of $N=2$ and $3$ in Figs. \ref{fig2}(b) and \ref{fig4}(c).}\label{fig5}
\end{figure}

As shown in Fig. \ref{fig5}, the laser beam can be aligned directly on the target atom in the plane formed by the atoms or in a slightly off-plane angle. This is sufficient to fulfill the asymmetric driving condition, where the spectator atoms serve as mediators that exchange photons with the target atom via the photon-mediated DDIs. As $N$ increases, we find a moderate enhancement of cooling performance as $\langle n_1 \rangle_{\rm multi} / \langle n_1 \rangle_{\rm single}\approx 0.84$ for $N=5$ comparing the cases of $\langle n_1 \rangle_{\rm multi} / \langle n_1 \rangle_{\rm single}\approx 0.95$ and $0.87$ for $N=2$ and $N=3$, respectively. However, we can also see that the cooling performance saturates as $N$ increases as well as even less enhanced cooling performance emerges in Fig. \ref{fig5}(e), which we attribute to the finite frequency shifts between the spectator atoms. The collective energy shift in the target atom from the spectator ones can not completely be avoided when even more atoms are added around it unless the atoms sit at the vertices of a hexagon. This suggests a weakening effect of multiatom enhancement in cooling, and a three-dimensional atomic structure can not resolve this issue since new projection angles between light polarizations and interparticle axes arise and lead to nonvanishing energy shifts on the target atom.   

\section{Discussion and Conclusion}

The enhanced dark-state sideband cooling investigated here can either utilize two-level or three-level atomic structures, where tunable external laser parameters can be tailored to fulfill the requirement of resolved sideband cooling condition. Therefore, our results can be applicable to subrecoil cooling of trapped atoms with surpassing performances by utilizing photon-mediated DDIs in atomic arrays with well-controlled positions. This outperformed cooling behavior results from long-range spin-exchange couplings which enable an enhanced cooling by removing extra heat under an asymmetric driving condition. Precise placements of atoms can be realized in versatile and reconfigurable optical tweezer arrays which have been applied in neutral atoms \cite{Barredo2016, Endres2016}, molecules \cite{Liu2018, Anderegg2019}, and ions \cite{Shen2020, Olsacher2020, Mazzanti2021}. With an exquisite control of atom arrays, the laser-cooling technique presented here is particularly useful and faster in preparing the motional ground state of atoms \cite{Kaufman2012, Thompson2013, Kaufman2021} without requiring atom-atom collisions in evaporative cooling. This presents the capability of long-range DDIs and their collective nature, which hosts notable spin-phonon correlations within the constituent atoms and opens up new parameter regimes for distinct cooling mechanisms. 

In conclusion, we theoretically investigate the dark-state sideband cooling of tapped atoms in free space with assistance of photon-mediated DDIs. We find that an enhanced cooling performance emerges and surpasses the conventional resolved sideband cooling in a single atom. This cooling mechanism is essential in driving the atom toward its motional ground state via spontaneous emissions. We present that a superior dark-state sideband cooling can be feasible when the atoms are placed at the magic interparticle distances, where the collective frequency shifts on the target atom becomes vanishing. An outperformed cooling behavior in the target atom also shows up in various multiatom configurations under varying laser detuning and light polarization angles, where we further demonstrate the multiatom improvement in superior cooling performance. Our results provide new insights to subrecoil cooling of atoms with assistance of collective DDIs in free space and pave the way toward realizing genuine quantum operations in multiple quantum registers.

\section*{ACKNOWLEDGMENTS}
We acknowledge support from the Ministry of Science and Technology (MOST), Taiwan, under the Grant No. MOST-109-2112-M-001-035-MY3. We are also grateful for support from TG 1.2 and TG 3.2 of NCTS at NTU and inspiring discussions with G.-D. Lin.

\end{document}